\documentstyle{mn}
\newread\epsffilein    
\newif\ifepsffileok    
\newif\ifepsfbbfound   
\newif\ifepsfverbose   
\newdimen\epsfxsize    
\newdimen\epsfysize    
\newdimen\epsftsize    
\newdimen\epsfrsize    
\newdimen\epsftmp      
\newdimen\pspoints     
\def\epsfbox#1{\global\def\epsfllx{72}\global\def\epsflly{72}%
   \global\def\epsfurx{540}\global\def\epsfury{720}%
   \def\lbracket{[}\def\testit{#1}\ifx\testit\lbracket
   \let\next=\epsfgetlitbb\else\let\next=\epsfnormal\fi\next{#1}}%
\def\epsfgetlitbb#1#2 #3 #4 #5]#6{\epsfgrab #2 #3 #4 #5 .\\%
   \epsfsetgraph{#6}}%
\def\epsfnormal#1{\epsfgetbb{#1}\epsfsetgraph{#1}}%
\def\epsfgetbb#1{%
\openin\epsffilein=#1
\ifeof\epsffilein\errmessage{I couldn't open #1, will ignore it}\else
   {\epsffileoktrue \chardef\other=12
    \def\do##1{\catcode`##1=\other}\dospecials \catcode`\ =10
    \loop
       \read\epsffilein to \epsffileline
       \ifeof\epsffilein\epsffileokfalse\else
          \expandafter\epsfaux\epsffileline:. \\%
       \fi
   \ifepsffileok\repeat
   \ifepsfbbfound\else
    \ifepsfverbose\message{No bounding box comment in #1; using defaults}\fi\fi
   }\closein\epsffilein\fi}%
\def\epsfsetgraph#1{%
   \epsfrsize=\epsfury\pspoints
   \advance\epsfrsize by-\epsflly\pspoints
   \epsftsize=\epsfurx\pspoints
   \advance\epsftsize by-\epsfllx\pspoints
   \epsfxsize\epsfsize\epsftsize\epsfrsize
   \ifnum\epsfxsize=0 \ifnum\epsfysize=0
      \epsfxsize=\epsftsize \epsfysize=\epsfrsize
     \else\epsftmp=\epsftsize \divide\epsftmp\epsfrsize
       \epsfxsize=\epsfysize \multiply\epsfxsize\epsftmp
       \multiply\epsftmp\epsfrsize \advance\epsftsize-\epsftmp
       \epsftmp=\epsfysize
       \loop \advance\epsftsize\epsftsize \divide\epsftmp 2
       \ifnum\epsftmp>0
          \ifnum\epsftsize<\epsfrsize\else
             \advance\epsftsize-\epsfrsize \advance\epsfxsize\epsftmp \fi
       \repeat
     \fi
   \else\epsftmp=\epsfrsize \divide\epsftmp\epsftsize
     \epsfysize=\epsfxsize \multiply\epsfysize\epsftmp   
     \multiply\epsftmp\epsftsize \advance\epsfrsize-\epsftmp
     \epsftmp=\epsfxsize
     \loop \advance\epsfrsize\epsfrsize \divide\epsftmp 2
     \ifnum\epsftmp>0
        \ifnum\epsfrsize<\epsftsize\else
           \advance\epsfrsize-\epsftsize \advance\epsfysize\epsftmp \fi
     \repeat     
   \fi
   \ifepsfverbose\message{#1: width=\the\epsfxsize, height=\the\epsfysize}\fi
   \epsftmp=10\epsfxsize \divide\epsftmp\pspoints
   \newcount\figskipcount
      \message{#1 \the\epsfysize  }
   \vbox to\epsfysize{\vfil\hbox to\epsfxsize{%
      \includegraphics{#1}%
      \hfil}}%
\epsfxsize=0pt\epsfysize=0pt}%

{\catcode`\%=12 \global\let\epsfpercent=
\long\def\epsfaux#1#2:#3\\{\ifx#1\epsfpercent
   \def\testit{#2}\ifx\testit\epsfbblit
      \epsfgrab #3 . . . \\%
      \epsffileokfalse
      \global\epsfbbfoundtrue
   \fi\else\ifx#1\par\else\epsffileokfalse\fi\fi}%
\def\epsfgrab #1 #2 #3 #4 #5\\{%
   \global\def\epsfllx{#1}\ifx\epsfllx\empty
      \epsfgrab #2 #3 #4 #5 .\\\else
   \global\def\epsflly{#2}%
   \global\def\epsfurx{#3}\global\def\epsfury{#4}\fi}%
\def\epsfsize#1#2{\epsfxsize}
\def\figinsert#1#2{\epsfbox{#1} \message{#2} }    
\def \Mpc {~h^{-1}~{\rm Mpc} }
\def \Om {\Omega_0}

\def \ro {r_0}
\def \bj {b_{\rm J}}
\def \ho {H_0 }
\def \ao {a_0 }

\def \kms {{\rm kms}^{-1}}

\title[Radio-quiet QSO environments - I.]
      {Radio-quiet QSO environments - I. The correlation of QSOs and
      $b_{\rm J}<23$ galaxies.}
\author[S.M. Croom \& T. Shanks]
       {Scott M.~Croom\thanks{S.M.Croom@durham.ac.uk} and
      T.~Shanks\\
Physics Department, University of Durham, South Road, Durham, DH1 3LE,
England.\\}
\begin{document}
 
\maketitle
\begin{abstract}
In this paper we present results of an analysis of radio-quiet QSO 
environments.  The aim is to determine the relation between QSOs, galaxies
and the mass distribution as a function of redshift.  
We cross-correlate a sample of $\sim150$ QSOs
from optically and X-ray selected catalogues with faint, $\bj<23$,
galaxies.  These data allow us to probe the galaxy clustering
environment of QSOs out to $z\sim1.0-1.5$.  Far from giving a positive
correlation, at $z<1.5$ the QSO-galaxy cross-correlation function is
marginally negative, with $\omega(\theta<120'')=-0.027\pm0.020$.  Colour
information suggests that the anti-correlation is most significant
between the QSOs and the red galaxy population. 

We have constructed models to predict the QSO-galaxy
cross-correlation, using the known form of the galaxy $N(z)$ 
at $\bj<23$, and assuming a variety of clustering evolution rates.
Cases in which QSOs exist in rich
cluster environments are comfortably ruled out at more than $5\sigma$
and the results are more consistent 
with a `normal' galaxy environment for radio-quiet QSOs.  If the small
anti-correlation is interpreted as an effect of gravitational lensing,
this conclusion is not altered.  In this case, the data are only
$\sim1\sigma$ below the low clustering amplitude models, while the
high amplitude models are still comfortably rejected.  We therefore
conclude that these QSOs may not be much more highly biased than
optically selected galaxies.
\end{abstract}
 

\section{Introduction}

Relating QSOs to the underlying mass distribution and to other mass tracers
is a vital step in the process of using QSOs to
measure large-scale structure in the Universe.  In order to determine
cosmological quantities, such as the density parameter $\Om$, from the
evolution of large-scale structure we are required to know the
evolution of the {\it mass} distribution.  Therefore, knowledge of the
biasing of QSOs with respect to this mass distribution is of the
utmost importance, and the relationship between galaxies and QSOs is an
essential step in understanding this bias.  The local environments of
QSOs also give important clues to the physical processes which form
these objects and control their evolution.

Direct imaging of the regions around QSOs is the most obvious way to
investigate their environments.  Yee \& Green (1984) \nocite{yg84}
imaged $3'\times3'$ fields centred on individual QSOs over a wide
range in redshift, ($0.05<z<2.05$).  They found significant numbers of
excess galaxies (to a limiting magnitude of $\sim21$ in the Gunn r
band) associated with QSOs at $z<0.5$.  The luminosity distribution of
the excess galaxies strongly suggested that these QSOs are associated
with galaxies at the distance implied by their cosmological redshift.
Further
investigations \cite{yg87} showed that at $z\simeq0.6$ some radio-loud
QSOs were found in environments as rich as those of Abell class 1
clusters.  Significant evolution was also detected from $z=0.4$ to
$z=0.6$ in this radio-loud sample, with the QSO-galaxy covariance
amplitude increasing by a factor of $\sim3$ as redshift increases.  
An extension of this
work \cite{eyg91} found that radio-quiet QSOs by contrast exist in
significantly poorer environments than their radio-loud
counter-parts. The results concerning radio-loud QSOs have been
confirmed by several other independent studies (e.g. Hintzen et al.
1991)\nocite{hrv91} which 
found significant numbers of excess galaxies associated with QSOs
out to $z\simeq1.5$.  However, observations of radio-quiet QSOs have
found a range of  results generally consistent with these objects
existing in poorer environments, similar to that of an average galaxy.
Smith, Boyle \& Maddox (1995) \nocite{sbm95} find that the
cross-correlation function between low redshift ($z\leq0.3$) X-ray
selected QSOs and $\bj<20.5$ galaxies from the APM Galaxy Survey
\cite{msel90} is consistent with the auto-correlation function of the
APM Galaxy Survey, suggesting that this population of QSOs is
unbiased with respect to galaxies.  Extending this work, Smith
(1998)\nocite{smith98} finds that X-ray selected QSOs to $z=0.7$ exist
in poor environments.  Boyle \& Couch (1993)
\nocite{bc93} have found no excess galaxy population associated with
radio-quiet QSOs at $z\sim1$, while Hutchings, Crampton \& Johnson
(1995) \nocite{hcj95} find that a number (9 out of 14) of radio-quiet QSOs
exist in
compact groups of star forming galaxies at $z=1.1$.  Clearly
there is a difference between radio-loud and radio-quiet QSOs,
with radio-loud QSOs inhabiting richer environments.  Radio-quiet QSOs
appear to inhabit environments similar to those of normal galaxies,
although the Hutchings et al. result is somewhat discrepant from this
hypothesis.  In this paper we will further investigate the
problem of QSO environments with a large ($\sim150$) sample of
optically and X-ray selected QSOs covering a wide range in redshift.
The second paper in the series (Croom \& Shanks 1998, in preparation)
will use deeper imaging to constrain QSO environments at higher
redshift.  

In Section 2 we describe our data and the methods employed.  We
present our cross-correlation results in Section 3, and
compare them to model predictions in Section 4.  A discussion of our
results and our conclusions are contained in Sections 5 and 6.

\section{Data and Methods}\label{xcordata}

\begin{table}
\baselineskip=20pt
\begin{center}
\caption{Details of the AAT photographic plates used in our
QSO-galaxy cross-correlation}
\vspace{0.5cm}
\begin{tabular}{ccccc}
\hline 
Field & Plate No. & R.A.& Dec. & Emulsion/Filter \\ 
Name & & (B1950) & (B1950)&\\
\hline 
SGP2 & J2801 & 00 49 39.4 & --29 21 34 & IIIaJ/GG385  \\
SGP4 & J1888 & 00 54 48.1 & --27 54 45 & IIIaJ/GG385  \\
QSF3 & J2719 & 03 40 18.0 & --44 18 14 & IIIaJ/GG385 \\
F855 & J1834 & 10 43 37.9 & --00 04 48 & IIIaJ/GG385 \\
& R1835 & 10 43 37.9 & --00 04 48 & IIIaF/RG630 \\
F864 & J1836 & 13 41 14.0 & --00 00 29 & IIIaJ/GG385  \\
& R1837 & 13 41 14.0 & --00 00 29 & IIIaF/RG630  \\
\hline
\label{aatplatetab}
\end{tabular}
\end{center}
\end{table}

\begin{figure}
\centering 
\centerline{\epsfxsize=8.0truecm \figinsert{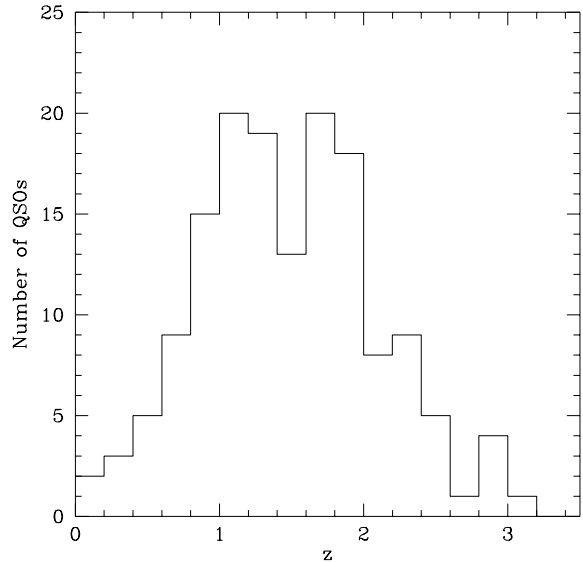}{0.0pt}}
\caption{The redshift distribution of all the QSOs, optically
and X-ray selected, used in this analysis.  The total number of QSOs is
152.}
\label{xoqsonz}
\end{figure}

The faint galaxy samples were taken from deep AAT plates in five
$40'\times40'$ fields.  Details of the plates are given in Table
\ref{aatplatetab}.  Each field has a $\bj$ plate and two fields
also have $r_{\rm F}$ plates.  The plates were scanned by the COSMOS
measuring machine. Details of the analysis of the first 3 fields
(SGP2, SGP4, QSF3), including star-galaxy separation  and photometric
calibration are given in Jones et al. (1991).\nocite{jfsep91}  Fields
F855 and F864 \cite{rsgsbg95,roche94} were
similarly analysed.  The plate scale for all five fields is 15.2
arcsec/mm.  The magnitude limit for the $\bj$ plates is
$\simeq24$ mag and the completeness limit is $\bj=23.5$.  We use the
galaxies to a limiting magnitude  of $\bj=23.0$, which is
comfortably brighter than the completeness limit.  At this magnitude
the errors in the star-galaxy separation are $\sim5\%$.

\begin{figure*}
\centering 
\centerline{\epsfxsize=18.0truecm \figinsert{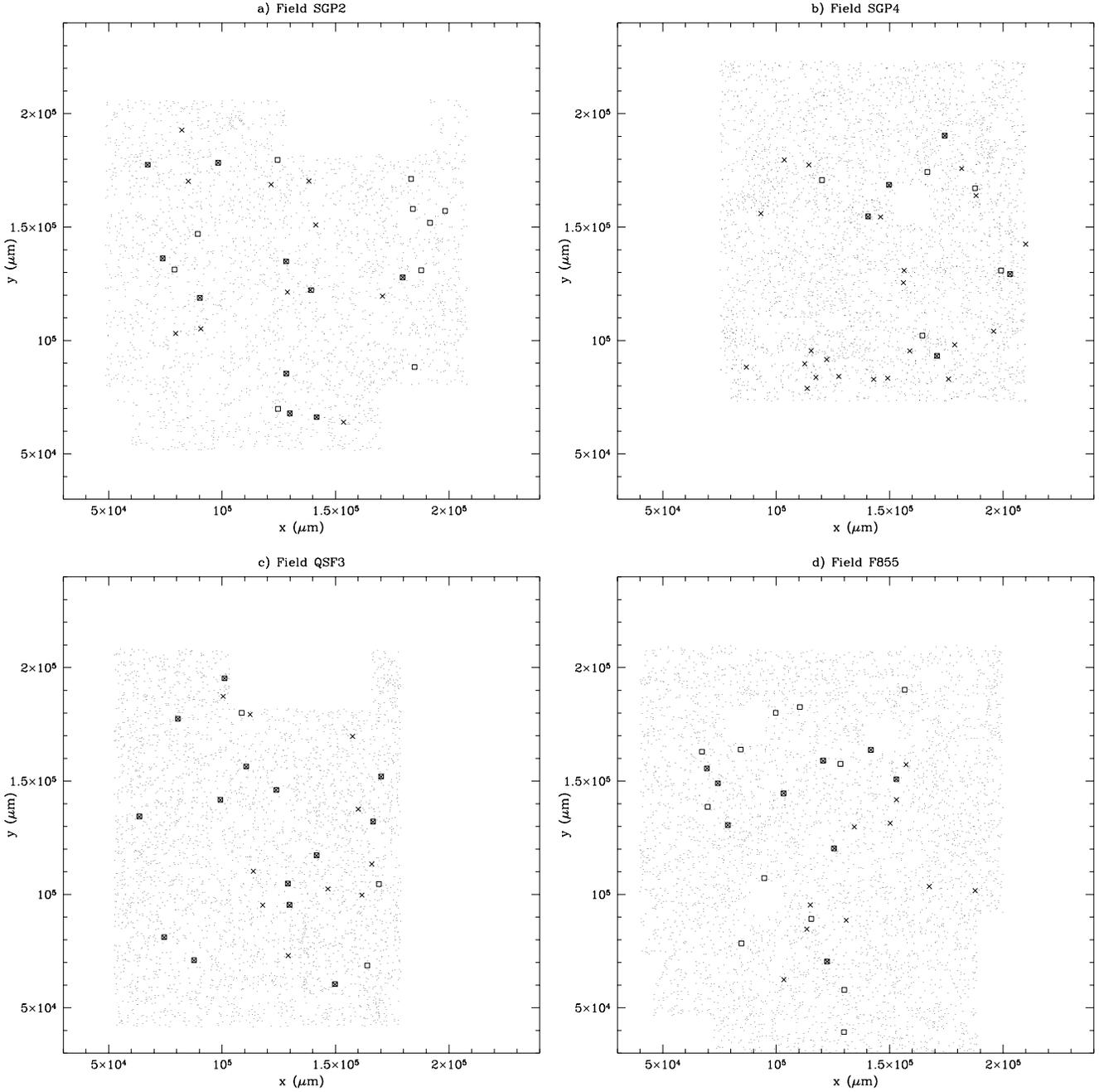}{0.0pt}}
\caption{The x,y positions of galaxies to $\bj=23$ from COSMOS
scans of AAT photographic plates, together with the positions of the
QSOs.  a) Field SGP2, b) field SGP4, c) field QSF3, d) field F855, e)
field F864. Open squares are the
optically selected QSOs while crosses are the X-ray selected QSOs,
The regions removed due to plate defects and bright stars can be
seen.  It should be noted that a number of QSOs are both optically and
x-ray selected.}
\label{aatqso}
\end{figure*}
\addtocounter{figure}{-1}

The QSOs in this analysis are from both optically and X-ray selected
samples.  Fields SGP2, SPG4 and QSF3 contain QSOs from the Durham/AAT
UVX selected sample \cite{bfsp90}, limited to $B=21$ mag.  The F855
and F864 fields contain QSOs from the deeper ESO/AAT colour selected
sample \cite{bjs91,zmzmb92}  with a limiting magnitude of $B=22.5$.
The X-ray selected QSOs were taken from optical follow-up of a deep
{\it ROSAT} survey \cite{omar96,rosat97}.  The QSOs are $>4\sigma$
detections with the {\it ROSAT} Position Sensitive Proportional Counter
(PSPC).  Details of the QSO samples used are given in Table
\ref{opxqsotab}, and the redshift distribution of the combined sample of
152 QSOs is shown in Fig. \ref{xoqsonz}.  It should be noted that a
large number of QSOs were selected independently using both optical and
X-ray techniques.  As pointed out in Section 1, radio-loud and
radio-quiet QSOs appear to have different environments.  We have
therefore attempted to determine whether any of our optically and X-ray
selected QSOs are, in fact, radio-loud.  Using the NRAO VLA Sky Survey
(NVSS) \cite{nvss} at $\delta>-40^{\circ}$ and the Parkes-MIT-NRAO (PMN)
Survey at $\delta<-40^{\circ}$ \cite{pmn} we find that none of our
QSOs are within $30''$ of a radio source at the $\sim2.5$mJy level
(NVSS) and $\sim25$mJy (PMN).  This is somewhat surprising, as we
expect $\sim5\%$ of optically selected QSOs to be radio-loud.

The QSO coordinates (optical counterpart positions in the case of the
X-ray selected objects) were transformed to the x,y system of the
COSMOS scanned $\bj$ plates using the Starlink ASTROM software;
$\sim40$ stars were used to produce the 6 coefficient transform.  The
RMS residuals for these transforms are shown in Table \ref{opxqsotab},
all are less than $1''$.  The five fields are plotted in Fig
\ref{aatqso}, showing all galaxies to $\bj=23$ with the
positions of the QSOs overlaid on top of this.

\begin{figure}
\centering 
\centerline{\epsfxsize=9.0truecm \figinsert{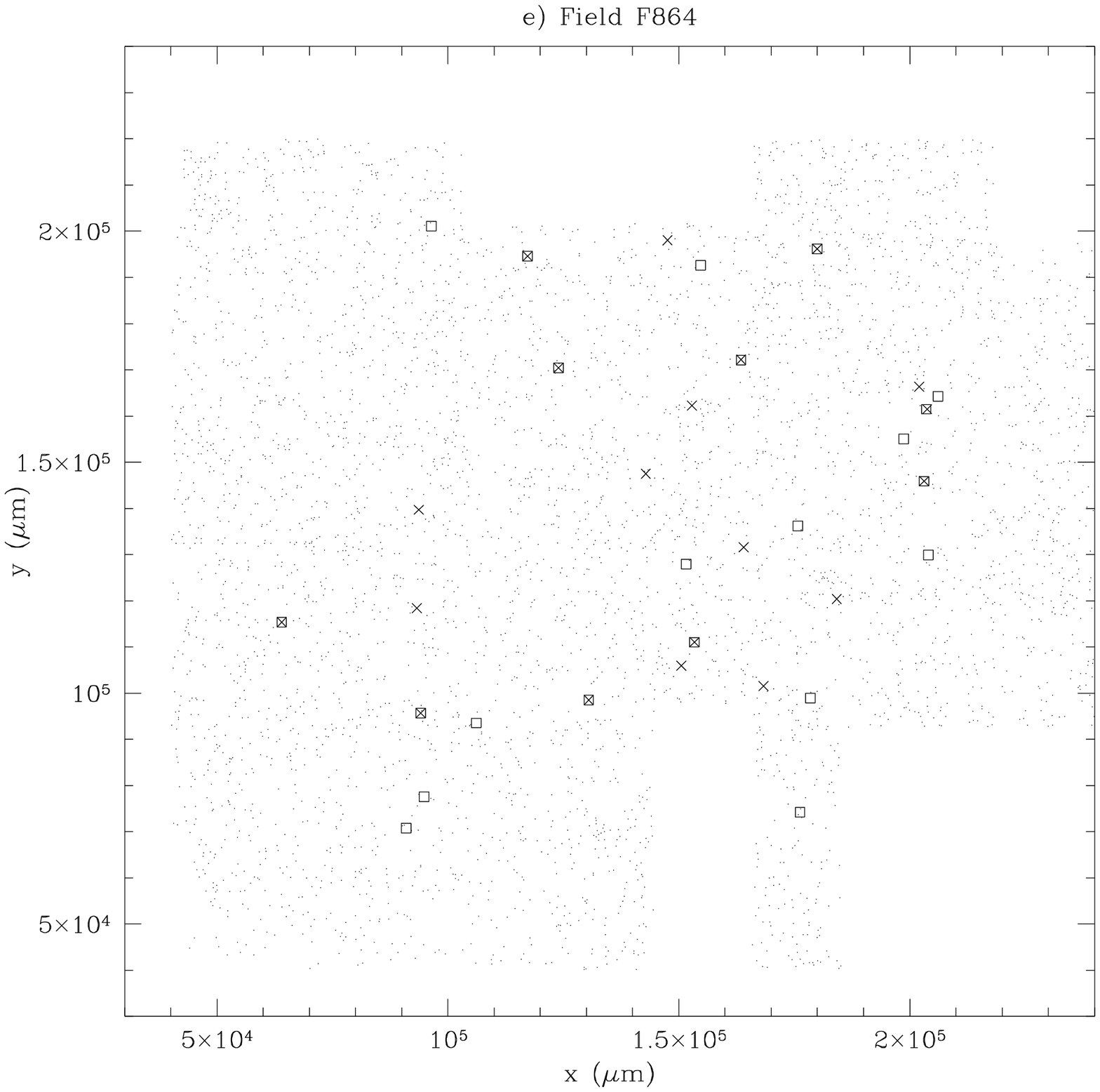}{0.0pt}}
\caption{Cont.}
\end{figure}

The cross-correlation was carried out in the standard manner, with the
galaxy distribution being compared to a random distribution of points
with the same angular selection function (that is avoiding holes) as
the galaxies.  The number density of randoms was 100 times the density
of galaxies.  The angular cross-correlation function, $\omega_{\rm
qg}(\theta)$, is defined as
\begin{equation}
\omega_{\rm qg}(\theta)=\frac{N_{\rm qg}(\theta)N_{\rm r}}{N_{\rm
qr}(\theta)N_{\rm g}}-1,
\label{omegathest}
\end{equation}
where $N_{\rm r}$ is the total number of random points, $N_{\rm g}$ is
the total number of galaxies, and $N_{\rm qg}$ and $N_{\rm qr}$ are
the number of QSO-galaxy and QSO-random pairs with separation
$\theta\pm\delta\theta/2$ respectively.  This was calculated in
concentric annuli of width $30''$.  The inner $5''$ is removed to
avoid any confusion between very near companions and the QSOs
themselves, and any QSOs which might have been classified as galaxies
on the COSMOS scans.  The errors shown in the figures below are
Poisson errors:
\begin{equation}
\Delta\omega=\frac{\omega(\theta)+1}{\sqrt{N_{\rm qg}}}.
\end{equation}
To test the Poisson error estimates we carry out two tests; the first
is to replace the QSOs with a random selection of galaxies and carry
out a cross-correlation between these random galaxies and the rest of
the galaxy sample.  A variance is calculated from 100 realizations of
this process.  The second test calculates the cross-correlation between
the QSOs and a bootstrap re-sampled galaxy distribution, again using
100 realizations to find a variance.  For all QSOs with
$z<1.5$ the Poisson error on the amplitude within $120''$ is
$\pm0.020$ (see Section \ref{xcorres} below).  The error estimated from
replacing QSOs with random galaxies is $\pm0.026$ while the bootstrap
estimator gives $\pm0.020$.  The excess variance from the first test
is due to the positive galaxy auto-correlation.  The bootstrap method
uses the QSO-galaxy cross-correlation which, as we show below, has no
significant clustering signal. 

\begin{table}
\baselineskip=20pt
\begin{center}
\caption{Details of the QSO samples used in this analysis.  It
should be noted that many QSOs were selected by both optical and X-ray
methods.}
\begin{tabular}{ccccc}
\hline 
Field & No. of & No. of & Total No. & r.m.s.\\
Name & Opt. QSOs & X-ray QSOs & of QSOs & Ast. ($''$)\\
\hline 
SGP2 & 20 & 20 & 30 & 0.86\\
SGP4 & 10 & 27 & 33 & 0.43\\
QSF3 & 17 & 24 & 26 & 0.39\\
F855 & 21 & 19 & 31 & 0.26\\
F864 & 22 & 20 & 32 & 0.69\\
\hline
\label{opxqsotab}
\end{tabular}
\end{center}
\end{table}

The background density of galaxies is determined from each field
separately which forces the integral of $\omega(\theta)$ to be zero
at the largest scales studied.  This is the well known `integral
constraint', which is simply due to the fact that there is still
appreciable clustering of galaxies on the scale of our fields.  In our
case we do not know {\it a priori} that there will be a significant
signal on scales of $\sim20'$, therefore an accurate estimate of the
integral constraint is not possible.  Given this, we can still
determine a possible integral constraint correction in order to
investigate its possible effect on our results.  The angular
correlation function determined in an area $\Omega$ is biased low by
the amount 
\begin{equation}
I=\frac{1}{\Omega^2}\int\int\omega(\theta_{\rm 12}){\rm d}\Omega_{\rm
1}{\rm d}\Omega_{\rm 2}. 
\end{equation}
We then assume the standard power-law for the cross-correlation function,
$\omega(\theta)=A\theta^{-0.8}$.  We find the following integral
constraint for the five fields:  $3.29A$ (SGP2), $3.42A$ (SGP4), 
$3.36A$ (QSF3), $3.04A$ (F855) and $2.89A$ (F864).  The
auto-correlation function amplitude for $\bj\sim23$ galaxies is
$\sim0.002$.  If we were to assume that this value was a reasonable
estimate of the QSO-galaxy cross-correlation then the integral
constraint in our fields would be $\sim0.006$.  We will demonstrate below
that the addition of an integral constraint of this amplitude has no
significant effect on our conclusions. 

\section{Cross-Correlation Results}\label{xcorres}

\begin{figure*}
\centering 
\centerline{\epsfxsize=16.0truecm \figinsert{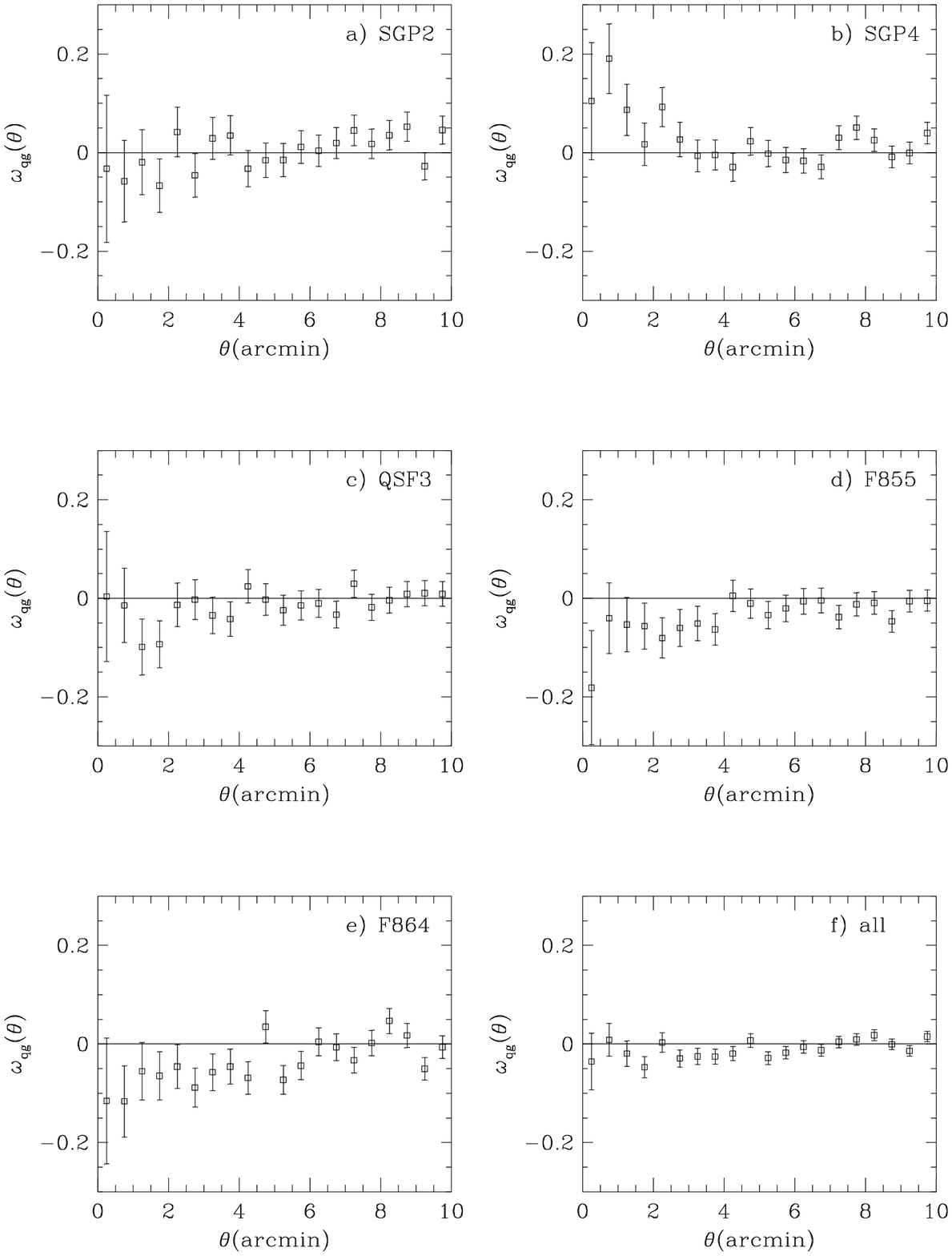}{0.0pt}}
\caption{The QSO-galaxy angular cross-correlation function for
each of the five AAT fields individually, including all QSOs.  f) shows
a pair weighted combination of all five fields.}
\label{qgwt}
\end{figure*}

We now present the results of the cross-correlation analysis between
both the optically and X-ray selected QSOs and the galaxy catalogues.
First, we look at the angular cross-correlation function for each
field individually;  these are shown in Fig. \ref{qgwt}a-e while  Fig.
\ref{qgwt}f shows a combination of all five fields.  We show the
results for all the known QSOs within each field (X-ray and optically
selected QSOs combined).  From these plots it is clear that
there is a significant {\it anti-correlation} in the F855 and F864 fields
while the SGP4 field is the only one which appears to show a positive
correlation.  The combined cross-correlation function therefore shows
a small negative signal on small scales.  Fig. \ref{qgwt} was derived
using all the QSOs in each field (including the whole range of
redshifts and both X-ray and optically selected objects).  In order to
look in detail at the source of the signal we calculate below a
clustering amplitude for each individual QSO.

\begin{figure*}
\centering 
\centerline{\epsfxsize=16.0truecm \figinsert{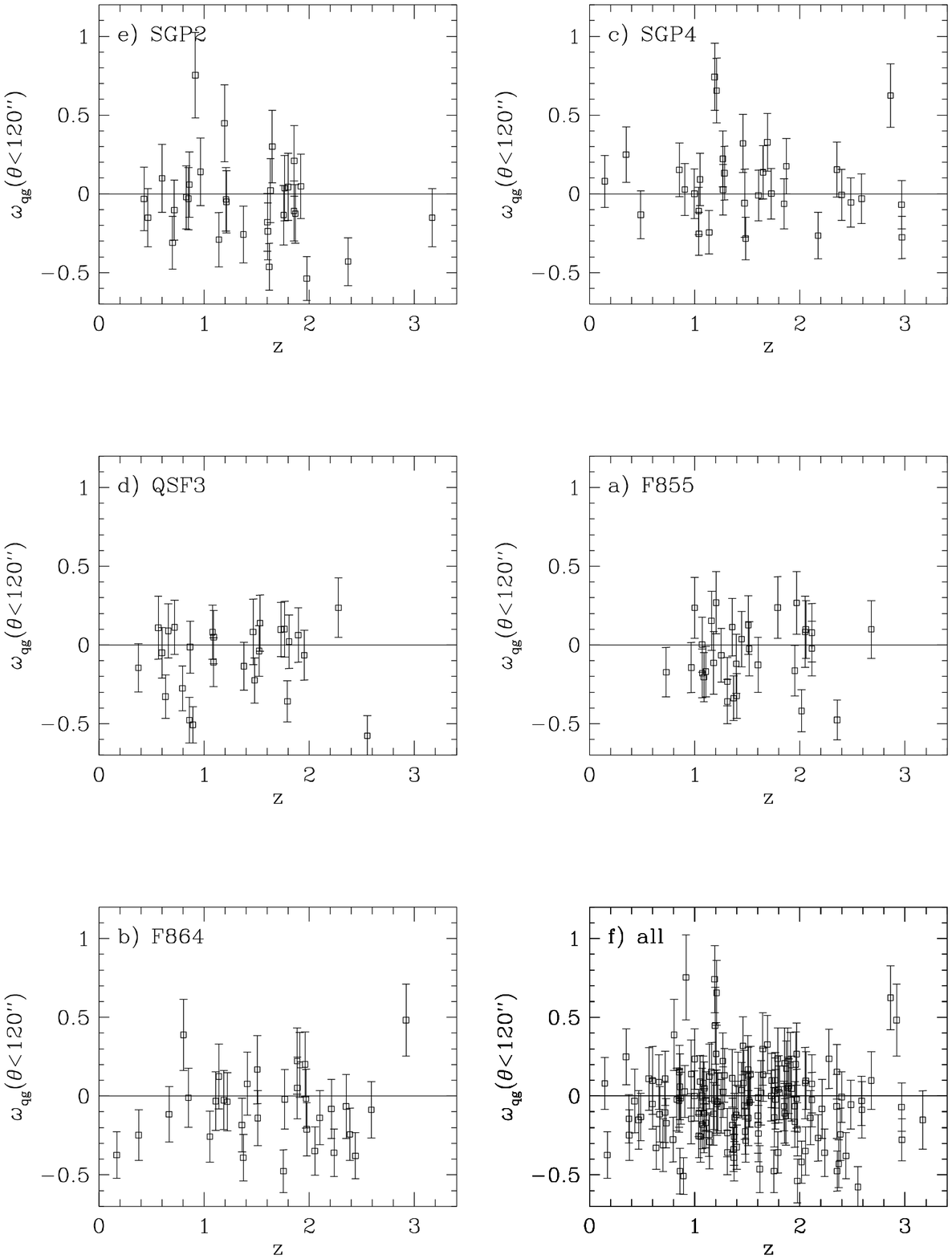}{0.0pt}}
\caption{The QSO-galaxy clustering amplitude for individual
QSOs in each AAT field (both optically and X-ray selected QSOs) as a
function of redshift.  f) shows all fields combined.  Individual
clustering amplitudes are measured within a radius of $120''$.}
\label{sixz}
\end{figure*}

\begin{figure*}
\centering 
\centerline{\epsfxsize=18.0truecm \figinsert{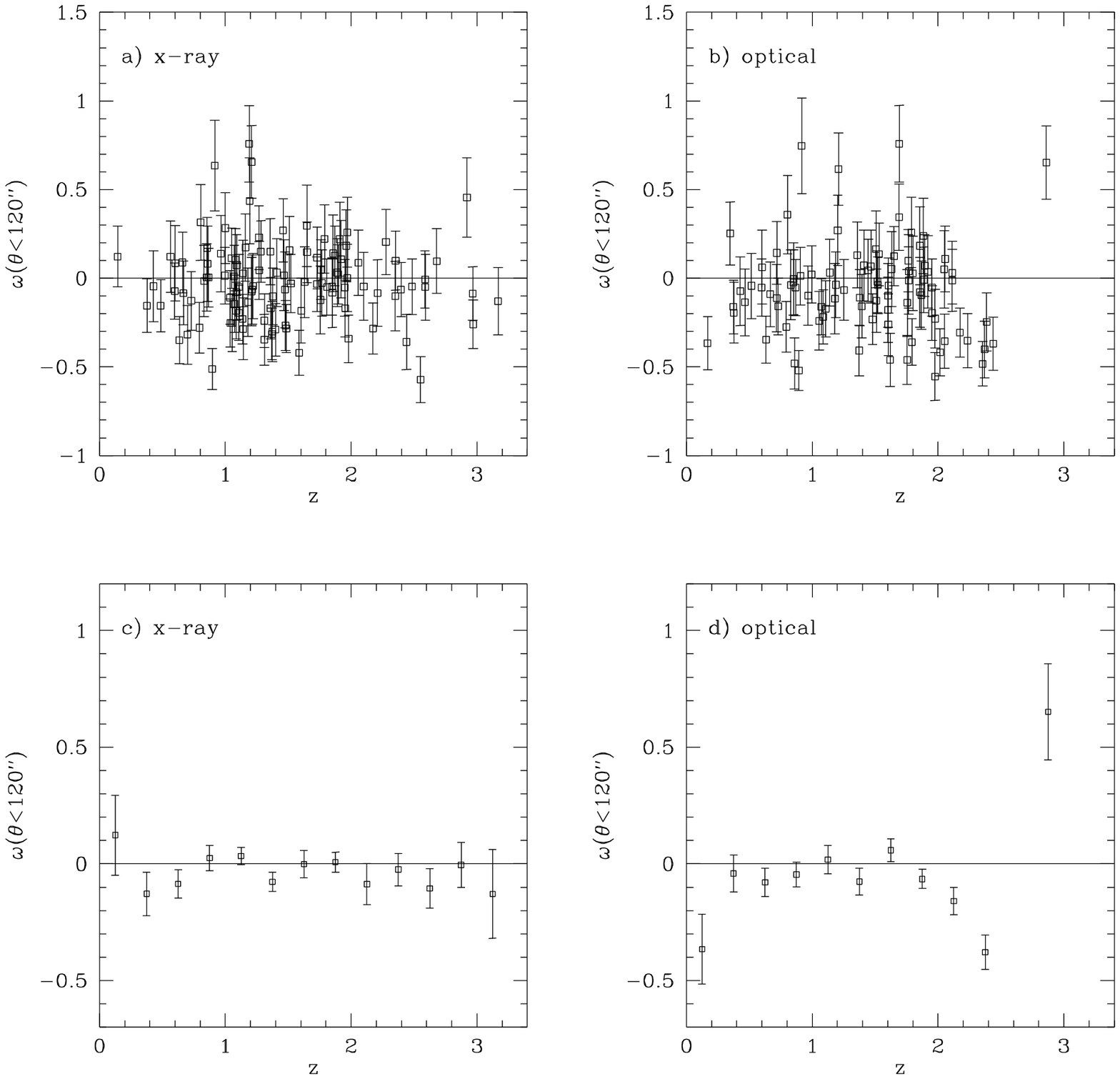}{0.0pt}}
\caption{QSO-galaxy clustering amplitudes within $120''$ for
QSOs as a function of redshift.  a) individual X-ray selected QSOs and
b) individual optically selected QSOs.  c) and d) show the same data
binned as a function of redshift with $\Delta z=0.25$.}
\label{ampz}
\end{figure*}

We calculate the individual cross-clustering amplitude for each QSO in
a non-parametric fashion, simply using the integrated excess of
galaxies out to a defined radius which we choose to be $120''$,
giving $\omega(\theta<120'')$ (still using an inner limit of $5''$).
A radius of $120''$ is equivalent to $0.5\Mpc$ in proper co-ordinates
at a redshift of $z\sim1$ for an $\Om=1$ flat Universe.  We can relate
this measure of clustering
to a parametric clustering amplitude assuming the simple power-law
form for the angular cross-correlation function, 
\begin{equation}
\omega_{\rm qg}(\theta)=A_{\rm qg}\theta^{1-\gamma}.
\end{equation}
The number of QSO-galaxy pairs between $\theta_{\rm 1}$ and
$\theta_{\rm 2}$ is
\begin{equation}
N_{qg}(\theta_{\rm 1}\leq\theta\leq\theta_{\rm 2})=\frac{2N_{\rm
g}}{\theta_{\rm 2}^2-\theta_{\rm 1}^2}\int_{\theta_{\rm 1}}^{\theta_{\rm 2}}
\theta(1+\omega_{qg}(\theta)){\rm d}\theta,
\end{equation}
where $N_{\rm g}$ is the number of galaxies.  This can then
be used to relate the non-parametric $\omega(\theta<120'')$ to the
clustering amplitude $A_{\rm qg}$ via Eq. \ref{omegathest} for
$5''\leq\theta\leq120''$.  We assume a power-law slope with
$\gamma=1.8$. 

In Fig. \ref{sixz} we show the clustering amplitudes for individual
QSOs within $120''$ as a function of redshift for each field.
These show that the positive correlation in the SGP4 field is mainly
due to three QSOs with strong positive signals.  One of these is at
$z=2.861$ and has $\omega(\theta<120'')=0.62\pm0.20$ 
(equivalent to $A_{\rm qg}=0.025\pm0.008$).  We do not expect
to see any galaxies brighter than $\bj=23$ at this redshift and hence
this is either a chance alignment or the effect of gravitational
lensing on the QSO.  Examination of this object by eye confirms that
there is a concentration of faint objects close to the QSO which
could be a cluster of galaxies.  The second of the three objects is only
separated from the $z=2.861$ QSO by $\sim1'$.  The third also shows a
number of faint objects very close to the QSO.    
By contrast, the negative signal in fields F855 and F864 is seen to be
due to a large number of objects with $\omega(\theta<120'')\sim-0.2$
to $-0.5$.  There is no measurable correlation between redshift and
clustering amplitude (determined from a non-parametric
rank-correlation test), with objects over the entire range in redshift
contributing to the negative signal.

  We should note
that using galaxies from the AAT plates to the plate limit of
$\bj\sim24$ does not have any effect on the results presented here.
The clustering amplitudes are not significantly altered, although
obviously the Poisson errors are somewhat reduced by the increased
number of galaxies.  Considering that at this magnitude limit the
sample is incomplete and the star-galaxy separation is ill-defined, a
limit of $\bj=23$ affords a more robust and reliable result.

\begin{figure*}
\centering 
\centerline{\epsfxsize=16.0truecm \figinsert{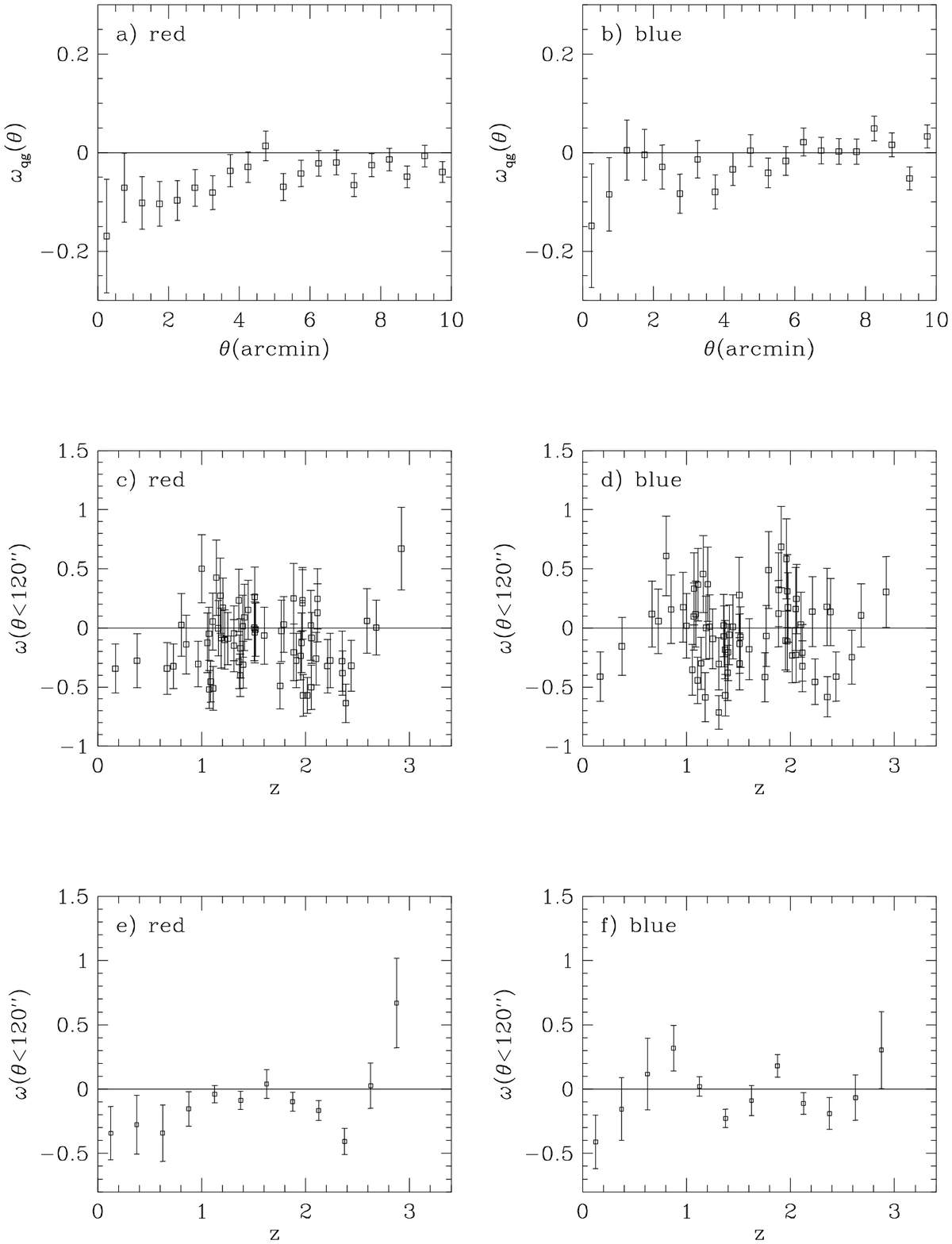}{0.0pt}}
\caption{Results for the QSO cross-correlation with colour
selected galaxies.  The angular cross-correlation
function for  a) red 
and b) blue galaxies. The clustering amplitude within $120''$ for
individual QSOs with c) red and d) blue galaxies as a function of
redshift, and the redshift binned clustering amplitude for e) red and
f) blue galaxies.}\label{colplots}
\end{figure*}

The clustering amplitudes for individual QSOs separated on the basis
of their selection (X-ray or optical) are shown in
Fig. \ref{ampz}.  It should be noted that a large
fraction of the QSOs were selected independently by both X-ray and
optical techniques.  Of course, the X-ray selected sample also has an
implicit optical selection imposed on it due to the need to identify
an optical counterpart to the X-ray source on which to carry out
follow-up spectroscopy.  Fig. \ref{ampz} shows that almost all of the
QSOs show zero or negative correlation amplitudes, while there are a
small number of outlying objects with significant positive signal.
The only difference between the X-ray
and optically selected QSOs is a possible anti-correlation in the
optical sample at high, $z>1.8$, redshift.  This can be seen more clearly
if the results are combined in redshift intervals, as shown in
Figs. \ref{ampz}c and d.  Here we combine the galaxy counts around each QSO
in $\Delta z=0.25$ intervals.  The X-ray selected sample
(Fig. \ref{ampz}c) shows no significant signal or trend as a function
of redshift.  The optically selected sample (Fig. \ref{ampz}d) shows
a significant anti-correlation at redshifts greater than $z\sim1.8$, with the
other significant points (in the first and last redshift bins)
both due to individual QSOs.  At low redshift, where any signal due to
real associations would be expected, none is detected; at $z<1.5$,
$\omega(\theta<120'')=-0.018\pm0.022$ for the X-ray selected sample
and $-0.051\pm0.027$ for the optical sample.  For the combined sample
of all QSOs with $z<1.5$ we obtain
$\omega(\theta<120'')=-0.027\pm0.020$ (82 QSOs).  In the next section these
results are compared to the expected values for given clustering
amplitudes and clustering evolution.

We have colour information in two fields, with F855 and
F864 containing $r_{\rm F}$ plates.  The colour distribution of the
galaxies peaks at
$\bj-r_{\rm F}\simeq1.25$ so we use this limit to define red and blue
galaxy populations.  Fig. \ref{colplots} shows the results of this
analysis, with Figs. \ref{colplots}a and \ref{colplots}b showing  the
angular 
correlation functions for all QSOs.  The red galaxies appear to
contribute more to the anti-correlation seen in the F855 and F864
fields.  When we look at the clustering amplitudes of individual QSOs
at $\theta<120''$ we see no obvious trend (Figs. \ref{colplots}c, d),
but binning these results as a function of redshift we see that the
blue galaxies show no strong correlation or anti-correlation at any
redshift, while the red galaxies appear to show generally negative
correlations (similar to that seen in the correlation of all galaxies
with the optically selected QSOs, Fig. \ref{ampz}d).  At $z<1.5$ the
correlations for the red and blue populations are
$\omega(\theta<120'')=-0.093\pm0.043$ and $-0.057\pm0.048$
respectively.  Further dividing the sample into X-ray and optically
selected QSOs we see no significant differences, this is mostly due to
the increased errors from a smaller number of QSOs.

We therefore see tentative evidence for a difference in the
cross-clustering properties of red and blue galaxies with QSOs.  Any
physical process which would cause this effect will be largely due to
the fact that red and blue galaxies tend to inhabit
different environments.  Red galaxies are found in richer
environments and are more strongly clustered than blue galaxies,
which tend to inhabit the field rather than clusters.  This
hints at gravitational lensing as a mechanism for the
anti-correlation.  Below we make
some estimates of the expected QSO-galaxy cross-correlation and relate
them to the results found here.

\section{Modelling QSO Environments}

\subsection{Limber's equation}

The angular cross-correlation function, $\omega_{\rm ab}(\theta)$,
between two populations, $a$ and $b$, can be related to the spatial
cross-correlation function $\xi_{\rm ab}(r,z)$ by Limber's equation
\cite{l53,peeb80} which can be written as
\begin{equation}
\omega_{\rm
ab}(\theta)=\frac{2\int_{0}^{\infty}\int_{0}^{\infty}x^4F^{-2}\phi_{\rm
a}(x)\phi_{\rm b}(x)\xi_{\rm ab}(r,z){\rm d}x{\rm
d}u}{\int_{0}^{\infty}x^2F^{-1}\phi_{\rm a}(x){\rm
d}x\int_{0}^{\infty}x^2F^{-1}\phi_{\rm b}(x){\rm d}x}, 
\end{equation} 
under the assumption that the integral is dominated by objects at
almost equal cosmic distance.  The $F$ term accounts for varying
cosmological geometries and is given by
\begin{equation}
F=\left[1-\left(\frac{\ho\ao x}{c}\right)^2(\Om-1)\right]^{1/2}.
\end{equation}
We will assume a flat cosmology with
$\Om=1$, that is $F=1$.  The comoving distance to a point midway
between $a$ and $b$ is given by $x$ (assuming that the separation of
$a$ and $b$ is small compared to $x$).  The small angle
approximation is also assumed so that the proper separation between
$a$ and $b$ is 
\begin{equation}
r=\frac{1}{1+z}\left(\frac{u^2}{F^2}+x^2\theta^2\right)^{1/2},\,\,\,\,\,\,\,\,\,\,\,\,\,\,\,\,\,\,u=x_{b}-x_{a}.
\end{equation}
$\phi_{\rm a}(x)$ and $\phi_{\rm b}(x)$ are the selection functions
for the two populations and satisfy 
\begin{equation}
\int_{0}^{\infty}x^2\phi(x){\rm d}x=\int_{0}^{\infty}N(z){\rm d}z,
\end{equation}
where $N(z){\rm d}z$ is the number of objects per unit surface area in
the redshift interval $[z,z+{\rm d}z]$.

We parameterize the spatial cross-correlation function in the
conventional power-law form:
\begin{equation}
\xi_{\rm
ab}(r,z)=\left(\frac{r}{\ro}\right)^{-\gamma}(1+z)^{-(3+\epsilon)},
\end{equation}
where $r$ is the proper separation of two objects and $\ro$ is the
correlation scale length at $z=0$.  Evolution is parameterized by
$\epsilon$, with various values of this parameter corresponding to the
following cases: $\epsilon=0$ is equivalent to constant clustering in
proper coordinates, so called {\it stable clustering};
$\epsilon=\gamma-3$ implies clustering which is constant in comoving
coordinates; $\epsilon=\gamma-1$ implies clustering which grows
according to linear theory (for $\Om=1$).  We adopt the standard
locally measured galaxy auto-correlation power-law slope of
$\gamma=1.8$.  Although this is not necessarily correct at high
redshift or in the case of QSO-galaxy cross-correlations, the
measured QSO auto-correlation is consistent with this slope \cite{cs96}.
We investigated the effect of altering the slope
and found that a change in slope of $\sim0.2$ does not significantly
alter our conclusions.

\begin{figure}
\centering 
\centerline{\epsfxsize=8.0truecm \figinsert{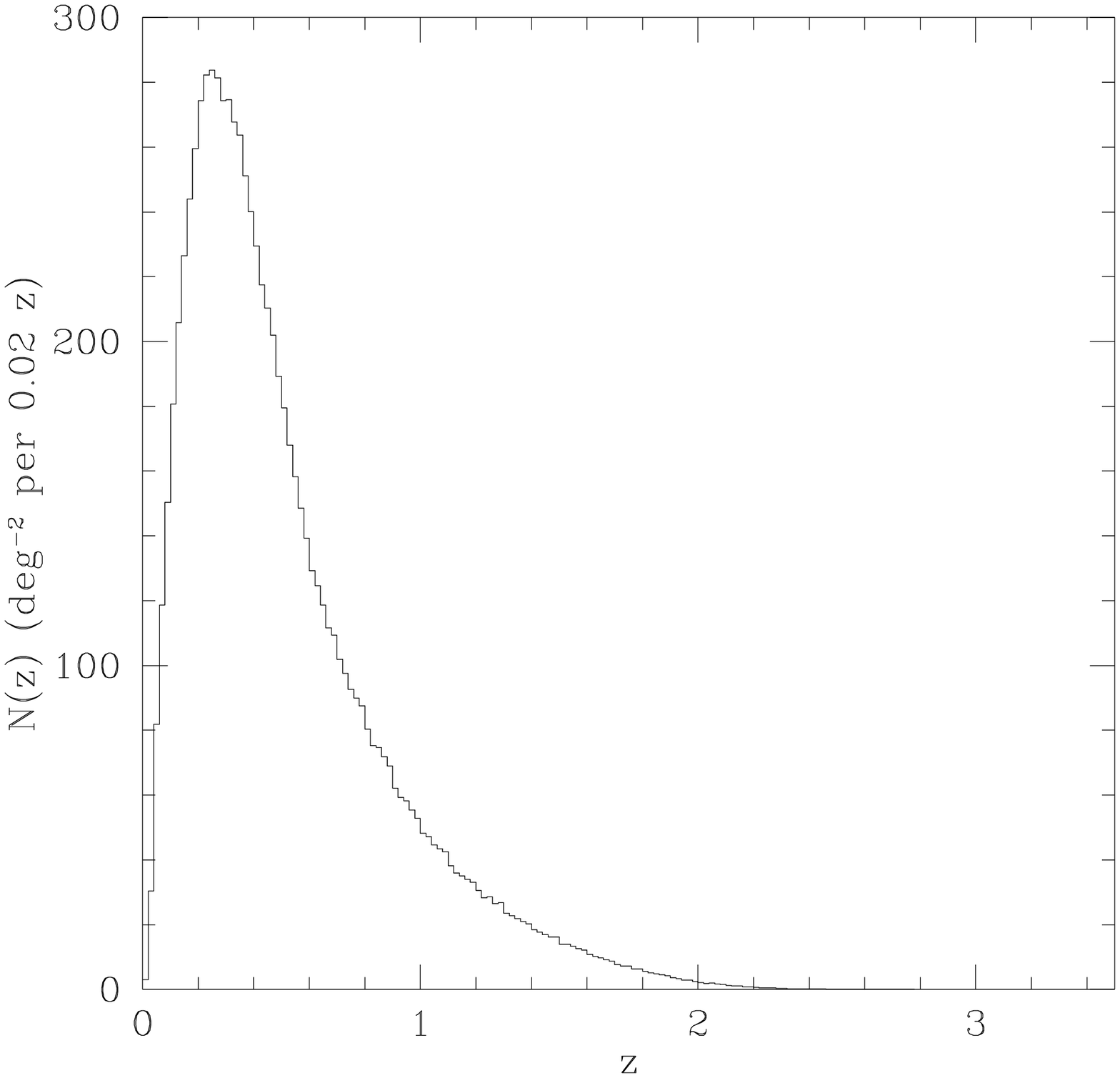}{0.0pt}}
\caption{The model redshift distribution from Metcalfe et al.
(1996) for a limiting magnitude of $\bj=23$.}
\label{galnz}
\end{figure}

The conversion from spatial to angular cross-correlation function is
obviously strongly dependent on the radial selection functions,
$\phi(x)$, of the two populations.  For the QSOs we simply use
the redshift distribution of the QSOs used in our analysis above
(Section \ref{xcorres}), that is, we integrate Limber's equation over
all the QSOs in the sample.  To model the galaxy selection function, we use
the $N(z)$ derived from the galaxy evolution model of Metcalfe et
al. (1996) \nocite{mscfg96} which is consistent with the redshift
distribution of galaxies in the deep Keck Telescope galaxy redshift
survey to $B=24$ \cite{chs95}.  These models contain an exponential
rise in star formation with look-back time and use the evolutionary
tracks of Bruzual \& Charlot (1993)\nocite{bruz93}.  The $N(z)$
distribution from this model is shown in
Fig. \ref{galnz}.  It should be noted that the luminosity evolution
produces a considerable high redshift tail to the distribution, as
found in the sample of Cowie et al., where significant numbers
of galaxies were found above $z=1$. 

Therefore, the only free parameters we have are the clustering
correlation length $\ro$ and the evolutionary parameter $\epsilon$,
which we attempt to constrain from the QSO-galaxy cross-correlation
results. 

\subsection{A comparison of models and data}\label{qgmodcomp}

\begin{table}
\baselineskip=20pt
\begin{center}
\caption{Predicted clustering amplitudes, $\omega(\theta<120'')$ and
$A_{\rm qg}$ (assuming the given value of $\gamma$), for $z<1.5$ QSOs
and $\bj<23$ galaxies in a number of different models.  $\sigma$ gives
the significance of 
the difference between the models and the observed clustering for all
QSOs with $z<1.5$.  No correction is made for the integral
constraint.}
\vspace{0.5cm}
\begin{tabular}{cccccc}
\hline 
$\ro$ & $\gamma$ & $\epsilon$ & $\omega(\theta<120'')$ & $A_{\rm qg}$ &
$\sigma$\\
\hline 
6.0 & 1.8 & -1.2 & 0.0430 & 0.00170 & 3.5\\
6.0 & 1.8 & 0.0 & 0.0245 & 0.00099 & 2.6\\
6.0 & 1.8 & 0.8 & 0.0175 & 0.00071 & 2.2\\
14.0 & 1.8 & -1.2 & 0.1976 & 0.00797 & 11.2\\
14.0 & 1.8 & 0.0 & 0.1128 & 0.00454 & 7.0\\
14.0 & 1.8 & 0.8 & 0.0805 & 0.00324 & 5.4\\
8.8 & 2.2& -1.2 & 0.3179 & 0.00233 & 17.2\\
8.8 & 2.2& 0.0 & 0.1867 & 0.00136 & 10.7\\
8.8 & 2.2& 0.8 & 0.1360 & 0.00100 & 8.2\\
2.0 & 1.8 & -1.2 & 0.0060 & 0.00024 & 1.7\\
2.0 & 1.8 & 0.0 & 0.0034 & 0.00014 & 1.5\\
2.0 & 1.8 & 0.8 & 0.0024 & 0.00010 & 1.5\\
\hline
\label{modres}
\end{tabular}
\end{center}
\end{table}

\begin{figure}
\centering 
\centerline{\epsfxsize=9.0truecm \figinsert{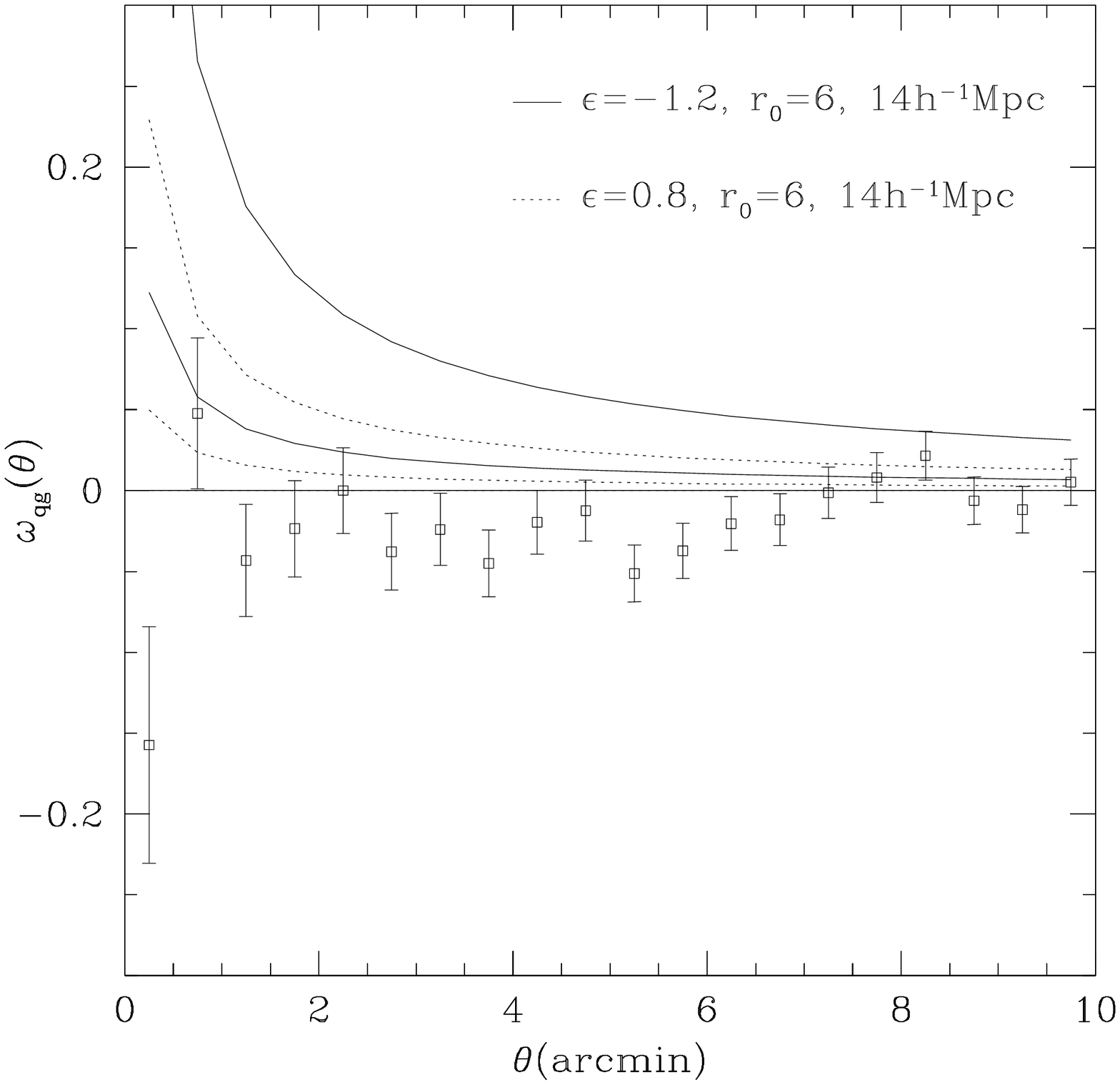}{0.0pt}}
\caption{$\omega(\theta)$ calculated from the models discussed
in the text compared to the results from all QSOs with $z<1.5$
correlated with $\bj<23$ galaxies.  The solid curves show models with
comoving clustering evolution ($\epsilon=-1.2$) and amplitudes of
$\ro=14\Mpc$ (upper) and $6\Mpc$ (lower).  The dotted curves show
models with linear theory clustering evolution ($\epsilon=0.8$) and
amplitudes of $\ro=14\Mpc$ (upper) and $6\Mpc$ (lower).}
\label{modcomp}
\end{figure}

\begin{figure}
\centering 
\centerline{\epsfxsize=9.0truecm \figinsert{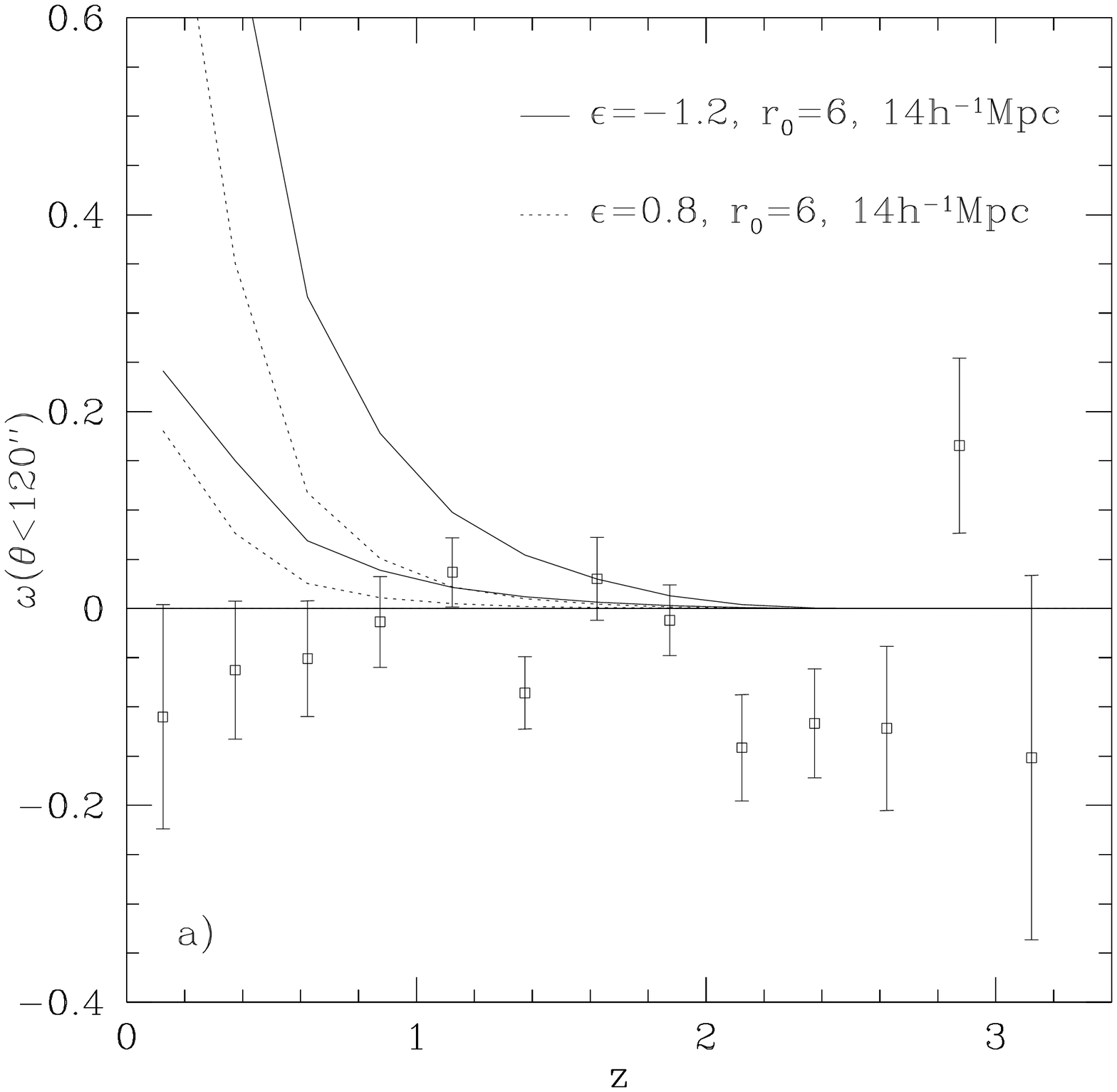}{0.0pt}}
\centerline{\epsfxsize=9.0truecm \figinsert{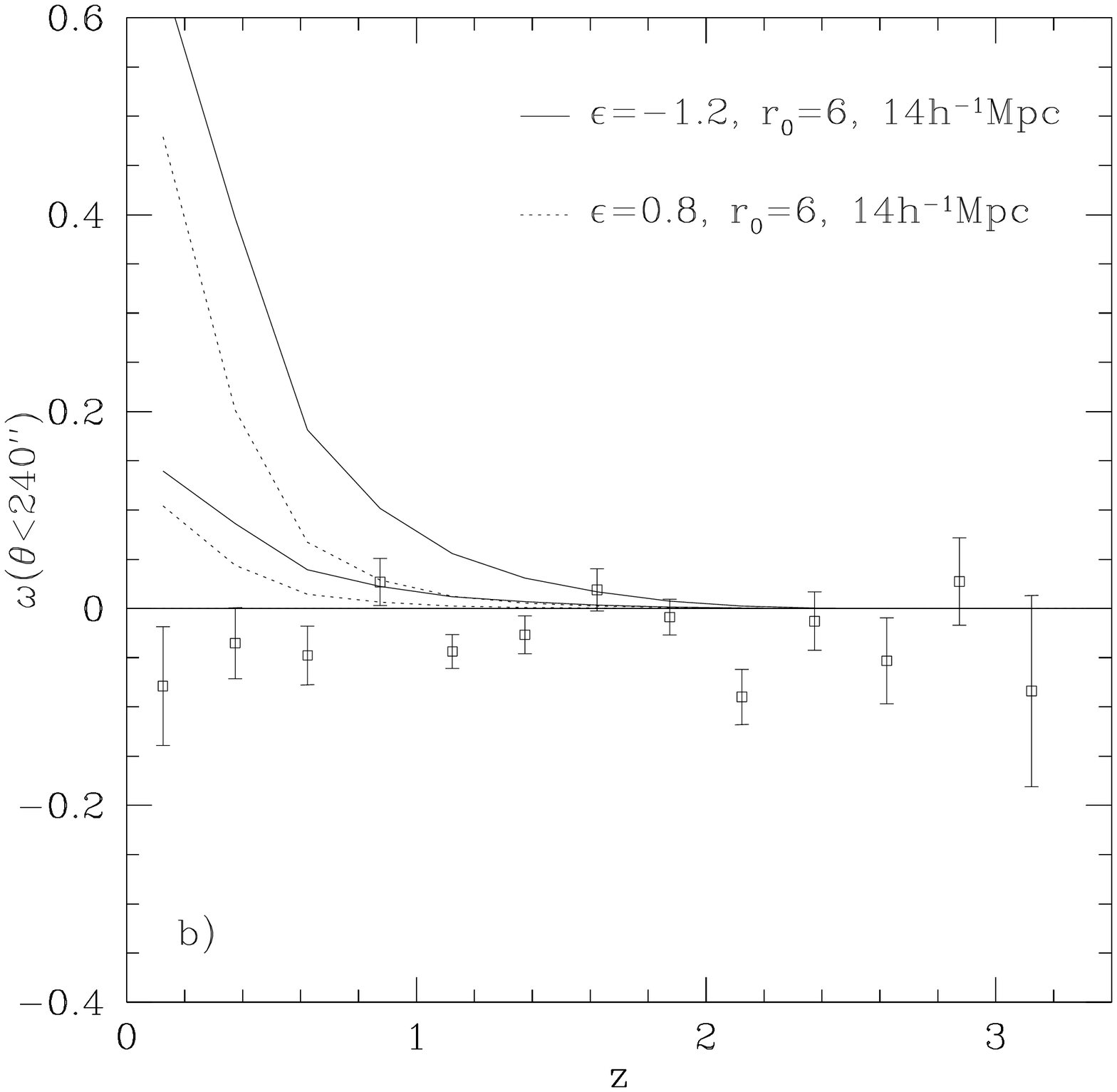}{0.0pt}}
\caption{a) $\omega(\theta<120'')$ and b) $\omega(\theta<240'')$
calculated from the models discussed in the text compared to the
results from all QSOs as a function of redshift.  The solid curves
show models with
comoving clustering evolution ($\epsilon=-1.2$) and amplitudes of
$\ro=14\Mpc$ (upper) and $6\Mpc$ (lower).  The dotted curves show
models with linear theory clustering evolution ($\epsilon=0.8$) and
amplitudes of $\ro=14\Mpc$ (upper) and $6\Mpc$ (lower).}
\label{modcomp2}
\end{figure}

We calculate the expected clustering amplitude,
$\omega(\theta<120'')$, for a number of
different cases which are shown in Table \ref{modres}.  We first
use the known present day galaxy-galaxy auto-correlation function with
$\ro=6\Mpc$ and $\gamma=1.8$.  With clustering evolution which
is stable in comoving coordinates this model is inconsistent with the
data for all $z<1.5$ QSOs at the $3.5\sigma$ level.  Clustering
which is stable in proper coordinates is rejected at the $2.6\sigma$
level and clustering evolution which is consistent with linear theory
is rejected at the $2.2\sigma$ level.  If we allow for an integral
constraint of $\sim0.006$, which of course assumes a clustering
amplitude similar to the auto-correlation amplitude of $\bj<23$
galaxies, the significance of the results are only changed by
$\sim0.3\sigma$.  This correction would be considerably smaller
if we used the measured cross-correlation amplitude
to estimate an integral constraint.  In fact the integral constraint
would be 
negative, increasing the significance of the above rejections.
Results from three other sets of parameters are also shown in Table
\ref{modres}.  With a clustering amplitude similar to that of the
cluster-cluster auto-correlation function $\ro=14\Mpc$ and $\gamma=1.8$
\cite{dalton94}, the lowest amplitude found (assuming linear theory
growth) is $5\sigma$ too high compared with the data.  We then try a model
which fits the galaxy-cluster cross-correlation function,
$\ro=8.8\Mpc$ and $\gamma=2.2$, \cite{le88}.  This model, which might 
be a better estimate of our expected signal if QSOs were found only in
rich clusters, is comfortably rejected for any reasonable rate of
evolution.  Finally we test a model with very low clustering
amplitude, $\ro=2\Mpc$, similar to that found in the Canada-France
Redshift Survey \cite{cfrs8} for $I_{\rm AB}<22.5$ galaxies with a
median redshift of 0.56.  This is clearly a better fit to our results,
being discrepant by only  $\sim1.5\sigma$.
 Ideally we would attempt to find the best fit values for
$\ro$ and $\epsilon$ using maximum likelihood techniques.  However, as
the measured cross-clustering amplitude is marginally negative, these
best fit values would have no physical meaning.  Fig. \ref{modcomp}
demonstrates that on scales $>120''$ the data are also
inconsistent with the models, out to $\sim8'$.  The cross-correlation
is, on average, negative up to this scale and inclusion of an integral
constraint of the magnitude discussed above (Section \ref{xcordata})
has no significant effect on this statement.

The models are also plotted as a function of redshift in
Fig. \ref{modcomp2}.  Fig. \ref{modcomp2}a is for
$\omega(\theta<120'')$ while Fig. \ref{modcomp2}b is for
$\omega(\theta<240'')$. These first demonstrate the relative
stability of using a $120''$ radius aperture to calculate the
clustering amplitude.  Secondly, they show the redshift at which the
models can rejected.  In Fig. \ref{modcomp2}b, all the models plotted
are inconsistent with the data for redshifts less than $z\sim1.5$.

\subsection{The effect of gravitational lensing}

Our results suggest that there could be a
lensing component to the cross-correlation of faint, $\bj<23$,
galaxies and optically selected QSOs, as a small anti-correlation was
measured between the two populations.  Similar anti-correlations
between brighter galaxies and UVX QSO candidates have been interpreted in
terms of gravitational lensing by Croom (1997).  This anti-correlation
is due to the two competing effects of magnification and area
distortion.  If the number-magnitude relation of the lensed objects is
flat, there are only a small fraction of objects that are added to the
sample through amplification, so that the area distortion wins,
reducing the perceived number density.

\begin{figure}
\centering
\centerline{\epsfxsize=8.8truecm \figinsert{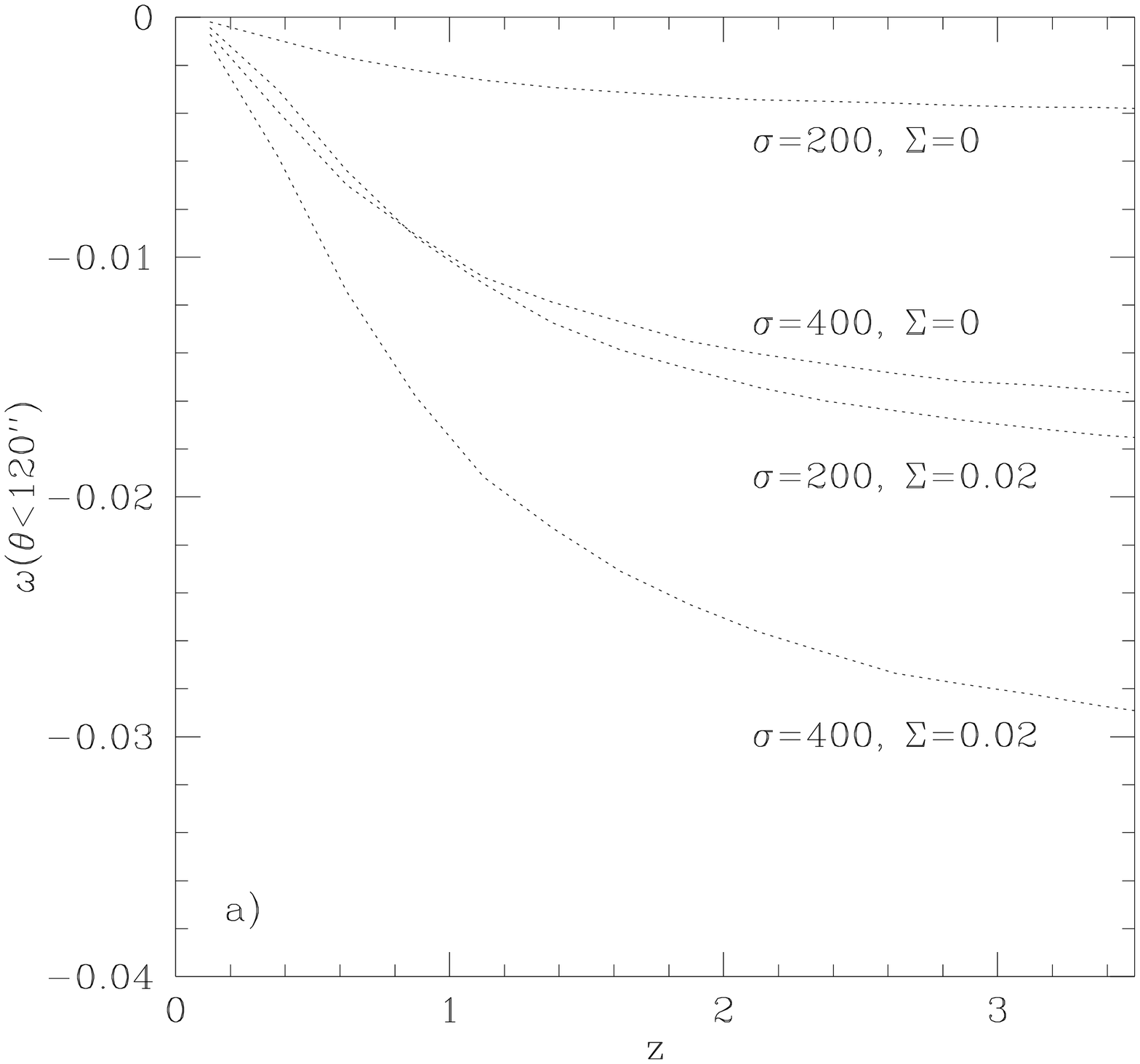}{0.0pt}}
\centerline{\epsfxsize=9.0truecm \figinsert{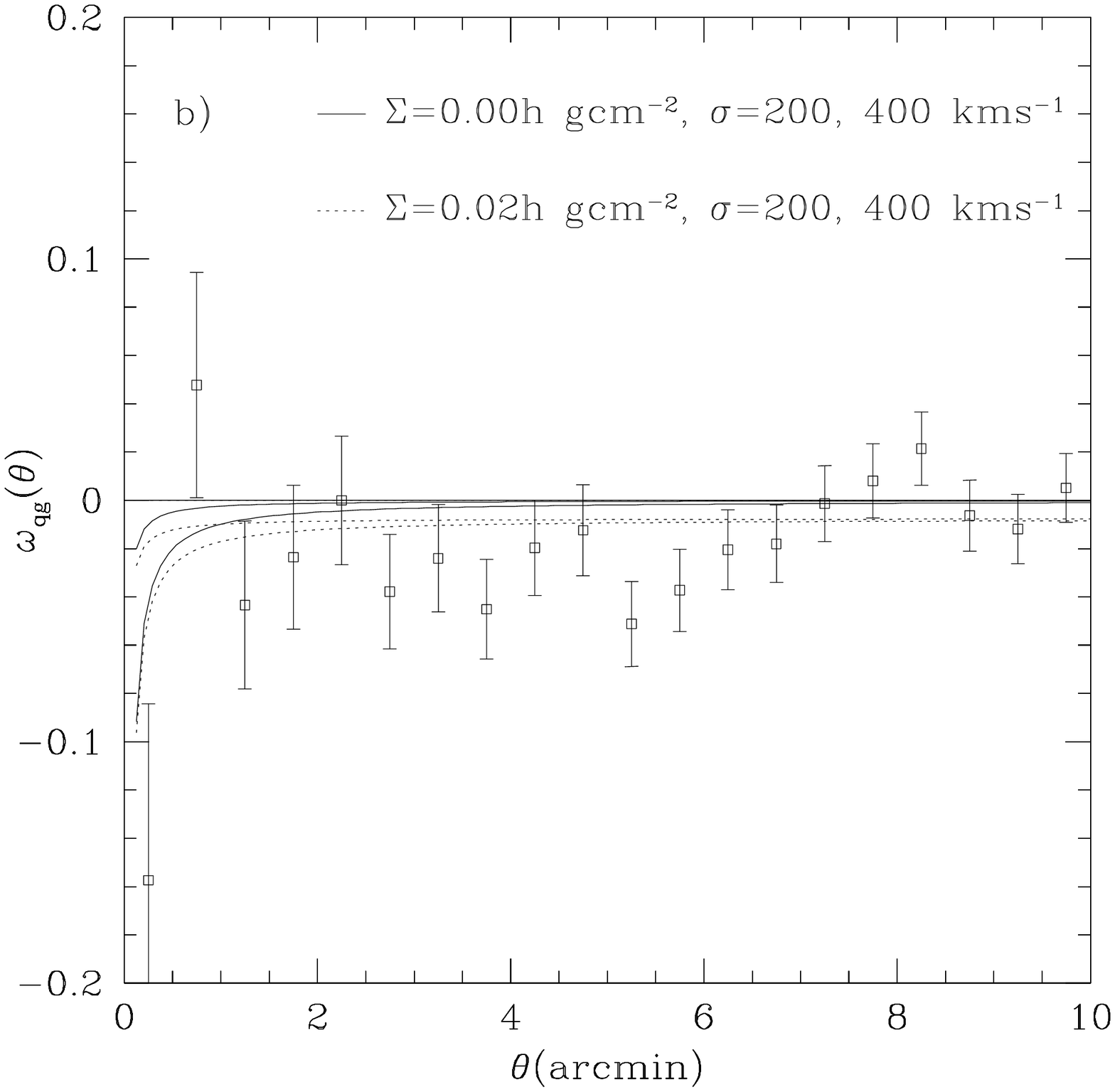}{0.0pt}}
\caption{Estimated angular cross-correlation function of QSOs
with faint, $\bj<23$, galaxies due to lensing. a)
$\omega(\theta<120'')$ as a function of QSO
redshift.  The correlation amplitude is measured within
$120''$ from Monte-Carlo simulations, with velocity dispersions $200$
and $400\kms$ and planes with $\Sigma=0.0$ and $0.02h~{\rm g~cm}^{-2}$mass
surface densities. b) $\omega(\theta)$ for the four different lensing models
for all QSOs with $z<1.5$ compared to the measured $\omega(\theta)$.}\label{monte} 
\end{figure}

Here we attempt to model the
effect of the lensing contribution on this result.  Using the redshift
distributions of the galaxy and QSO populations
we construct Monte-Carlo simulations to calculate the
expected correlation signal as a function of redshift.  These
simulations model the individual galaxy potentials as isothermal
spheres, and take into account the clustering of galaxies by also
adding a constant density plane.  Wu et al.
(1996)\nocite{wfzq96} calculate that the maximum matter surface
density contributed by large-scale structure is $\sim0.01-0.02h~{\rm
g~cm}^{-2}$ assuming that the amount of matter in galaxies is
$\Omega_{\rm g}\sim1$.  Typical velocity dispersions for galaxies are
$\sim200\kms$, so we produce Monte-Carlo simulations with velocity
dispersions of 200 and $400\kms$ and constant density planes with
$\Sigma=0.0$ and $0.02h~{\rm g~cm}^{-2}$.  We use a value of 0.28 for
the slope of the QSO integral number-count slope.  Fig. \ref{monte}
shows how 
the clustering amplitude in these simulations changes with redshift.
The extreme case of $\Sigma=0.02h~{\rm  g~cm}^{-2}$ and
$\sigma=400\kms$, which can contribute $\sim-0.02$ to the correlation
amplitude, can
account for most of the negative signal found for QSOs at $z<1.5$,
while it
cannot account for the negative correlation found at higher $z$.
Fig. \ref{monte}b shows the $\omega(\theta)$ predicted by the 
lensing models, for a combination of all the QSOs at $z<1.5$, compared
to the data.  Correcting the data to allow for this lensing effect
would not give the correlation function a positive amplitude.  Lensing
has no effect on the rejection of high clustering amplitude models,
but does make the models with low clustering
amplitude more consistent with the data.

\section{Discussion}

It is clear that there is strong disagreement between the clustering
results presented here and a range of models.  This is
demonstrated by Fig. \ref{modcomp}.  At $z<1.5$ the
correlation within $120''$ is $\omega(\theta<120'')=-0.027\pm0.020$,
and at $z>1.5$, 
the combined X-ray and optically selected sample has a clustering
amplitude of $\omega(\theta<120'')=-0.028\pm0.021$, almost identical
to that found at lower redshift.  If we assume that a non-zero
cross-correlation is only produced by physically associated QSOs and
galaxies then the above results give us strong constraints on the
environments of radio-quiet QSOs up to $z\sim1-1.5$.  Out to this
redshift we still expect to see significant numbers of galaxies with
a limiting magnitude of $\bj=23$.  The local luminosity function has
$M_{\rm J}^{*}\simeq-19.7$ (e.g. Ratcliffe 1996)\nocite{ratcliffe96}
which is equivalent to $\bj=23$ and $24$ for $z=1.0$ and $z=1.5$
without K-corrections.  A combination of K-correction and
reasonable luminosity evolution \cite{pbz96} leaves these values
largely unchanged (particularly for late type galaxies) and thus at
$z=1$ we reach $\sim M^{*}$ and at $z=1.5$ about 1 mag brighter than
$M^{*}$.  We can firmly reject the hypothesis that the QSOs in our
data exist in rich environments similar to galaxy
clusters.  This is true for a wide range of evolutionary rates
(parameterized by $\epsilon$).  A scenario in which QSOs have poorer
environments, similar to `normal' galaxies, is a more acceptable
alternative, although this still gives a clustering amplitude which is
inconsistent with the data at the $2-3\sigma$ level.

We see a significant anti-correlation in our data, which cannot be
explained by the conventional clustering models used above, but the data 
give us various clues as to the source of this signal.  First, we
find that the optically selected QSOs show more anti-correlation than
the X-ray selected samples, particularly at high redshift.  Secondly,
when we sub-divide the galaxy samples on the basis of colour (for only
two fields), we find that the red galaxies show a stronger
anti-correlation than the blue galaxies.  It has been established
that the auto-correlation amplitude for faint red galaxies is
significantly higher than for faint blue galaxies \cite{rsmf96}.  This is
easily understood if the redder galaxies lie predominantly at low
redshift while the high redshift tail of the galaxy distribution is
composed of the blue population.  Of course, early type galaxies are
also more numerous in rich structures such as galaxy clusters.  It
therefore appears that QSOs tend to show an anti-correlation with
galaxies in rich environments at low redshift.  

The possible causes of
this signal are either obscuration by high densities of dust in rich
structures or gravitational lensing which can cause an
anti-correlation if the QSO number-counts slope is flat (e.g. Narayan
1989)\nocite{n89}.  The optical QSOs in the sample are in the magnitude
range where the number counts are flat, $\sim0.3$, compared to the
critical slope of $0.4$ which is the slope required for
zero-correlation.  The X-ray 
selected QSOs, in contrast, have a ${\rm log}N-{\rm log}S$ slope of
$2.52\pm0.3$ at bright magnitudes turning over to $1.1\pm0.9$
\cite{gssbg96}.  The critical ${\rm log}N-{\rm log}S$ slope is $1.0$ so
we expect no anti-correlation (although it should be noted that the
errors in the slope at faint fluxes are large).  Above we show that
even extreme estimates of galaxy masses could
only produce anti-correlations of $\omega(\theta)\sim-0.02$ to
$-0.03$.  This extreme 
model can just account for the anti-correlation measured.
Lensing could reduce the inconsistency between our results
and the lowest amplitude clustering models to only $\sim1\sigma$,
while the inconsistency with the strongly clustered models is
still very significant.  Lensing does not appear to
explain the larger anti-correlation found at $z>1.8$ between the
$\bj<23$ galaxies and optical QSOs, although it should be noted that
this signal is due to a relatively small number of QSOs.
A final possibility is that
QSOs at low redshift might avoid rich environments.  However, we have
only a small number of QSOs at low redshift which are in fields that
contain colour information (7 QSOs at $z<1$), and also, other observations
\cite{sbm95} find a low but positive correlation between QSOs and
galaxies at low redshift.

Finally we note that deep wide field CCD imaging will allow this study
to be developed further.  Deep, $B<26$, imaging in an area of
$\sim2$ sq. deg. will be presented in the second paper in this
series (Croom \& Shanks 1998 in preparation).  This data will allow us
to probe QSO environments at higher redshift $z\sim2$, and determine
if the present study may have missed small, faint galaxies associated
with lower redshift QSOs.

\section{Conclusions}

We have carried out an investigation of the environments of a large
sample of QSOs ($\sim150$), using deep AAT photographic plates in 5
independent areas.  The QSOs were selected by both optical colour
techniques and deep X-ray exposures from {\it ROSAT}.  We draw
the following conclusions:

1. Most QSOs show zero or negative correlations.  For a combination of
all QSOs at $z<1.5$ we find $\omega(\theta<120'')=-0.027\pm0.020$.
Only a small number of QSOs ($\sim4-5$) show significant positive
correlation with $\bj\leq23$ galaxies; 

2.  Optically and X-ray selected QSOs show marginally different
properties, with the optical sample being more anti-correlated than
the X-ray sample; for $z<1.5$ the X-ray selected sample gives
$\omega(\theta<120'')=-0.018\pm0.022$, while the optically selected
sample gives $\omega(\theta<120'')=-0.051\pm0.027$.  At high redshift,
$z>2$, the optical QSOs become more anti-correlated with galaxies.

3.  Analysis of the two fields with colour information suggests that the
anti-correlation is stronger in the red galaxy population than in the
blue population.

4.  The cross-correlation results are compared with models which include
clustering evolution.  Models which have clustering amplitudes similar
to that of galaxy clusters ($\ro\sim 14\Mpc$) with a wide range of
evolutionary rates are ruled out.  The data are more consistent with a
low clustering amplitude, similar to that of the galaxy
auto-correlation function, although $\ro=6\Mpc$ combined with linear theory
evolution still gives too high an amplitude at the $2\sigma$ level.
Models with $\ro=2\Mpc$ are only $\sim1.5\sigma$ above the observed
cross-correlation.  If
gravitational lensing causes some of the measured anti-correlation,
the disagreement with the low amplitude clustering model could be
reduced to $\sim1\sigma$.

5.  We suggest that the anti-correlation found could be due to
gravitational lensing, particularly as the optical QSOs and red
galaxies show the strongest anti-correlation.
An extreme lensing model could account for most of the measured
anti-correlation at $z<1.5$.  However, even this extreme lensing model
does not significantly affect our conclusions when comparing the data to
clustering models. 

The results presented here clearly reject the notion that radio-quiet
QSOs exist in rich cluster-like environments up to $z\sim1-1.5$.  Beyond
this redshift we require deeper images to determine the environments
of QSOs, these observations will be important in the interpretation of
new QSO large-scale structure measurements from the next generation of
QSO surveys (e.g. the 2dF QSO Redshift Survey, Smith et al. 1997).

  Our results are consistent with those of Boyle \& Couch
(1993) who find no correlation between $R\sim23$ galaxies and QSOs in
the redshift range $0.9<z<1.5$.  Ellingson et al., (1991) find a
marginal positive signal at $z<0.6$, which is consistent with our data
(although we have a small number of QSOs at these low redshifts).  Our
data appear to be inconsistent with those of Hutchings et al., (1995)
who find that a sample of 9 out of 14 QSOs are associated with compact
groups of star forming galaxies, although these objects lie in a QSO
{\it super-cluster} and so may be a unrepresentative sample existing in a
particularly rich environment.  We conclude that our results continue to
uphold the hypothesis that QSOs are not strongly biased with respect
to the galaxy population, but are more likely to trace the
distribution of {\it normal} galaxies.

\section*{acknowledgements}

This paper was prepared using the facilities of the STARLINK node at
the University of Durham.  SMC acknowledges the support of a Durham
University Research Studentship.  We thank Robert Smith for many
useful comments concerning this paper.

{}
\end{document}